\documentclass[10pt, conference, compsocconf]{IEEEtran}
\IEEEoverridecommandlockouts                              
\usepackage{graphics,color} 
\usepackage{subfigure}
\usepackage{rotating}
\usepackage{algorithmic}
\usepackage{comment}
\usepackage{algorithm,amssymb,url}
\usepackage{epsfig}
\usepackage[cmex10]{amsmath}
\usepackage{amsthm}
\usepackage{bm}
\usepackage{cite}
\usepackage{textcomp}

\definecolor{orange}{rgb}{1,0.5,0}

\definecolor{lblue}{rgb}{0,0.4,0.8}

\newcommand {\eeq} {\end{equation}}
\newcommand {\barr} {\begin{array}}
\newcommand {\earr} {\end{array}}
\newcommand {\bear} {\begin{eqnarray}}

\newcommand {\namep} {Cloud4IoT\relax}
\newcommand {\Iotap} {IoT support application}
\newcommand {\Datap} {Data logic \& processing application}
\begin{document}
\title{\namep: a heterogeneous, distributed and autonomic cloud platform for the IoT\footnote{}}

\author{Daniele Pizzolli, Giuseppe Cossu, Daniele Santoro, Luca Capra, Corentin Dupont, \\
Dukas Charalampos, Francesco De Pellegrini, Fabio Antonelli and Silvio Cretti\\
CREATE-NET, via alla Cascata 56/D, 38123 Trento, Italy.
\thanks{The authors are with CREATE-NET, via Alla Cascata 56/D, 38123 Trento, Italy; email: \texttt{name.surname@create-net.org}. This research received funding from the European Union's H2020 Research and Innovation Action under grant agreement n. 688088 (project AGILE) and grant agreement n. 687607 (project WAZIUP).}}

\IEEEoverridecommandlockouts
\IEEEpubid{\makebox[\columnwidth]{\hfill
  9781-5090-1445-3/16\$31.00~\copyright~2016 IEEE (CloudCom'16)
  }
  \hspace{\columnsep}\makebox[\columnwidth]{}
}
\maketitle

\begin{abstract}
We introduce \namep, a platform offering automatic deployment, orchestration and dynamic configuration of IoT support software components and data-intensive applications for data processing and analytics, thus enabling plug-and-play integration of new sensor objects and dynamic workload scalability. \namep~enables the concept of {\em Infrastructure as Code} in the IoT context: it empowers IoT operations with the flexibility and elasticity of Cloud services. Furthermore it shifts traditionally centralized Cloud architectures towards a more distributed and decentralized computation paradigm, as required by IoT technologies, bridging the gap between Cloud Computing and IoT ecosystems. Thus, \namep~is playing a role similar to the one covered by solutions like Fog Computing, Cloudlets or Mobile Edge Cloud.

The hierarchical architecture of \namep~ hosts a central Cloud platform and multiple remote edge Cloud modules supporting dedicated devices, namely the IoT Gateways, through which new sensor objects are made accessible to the platform. Overall,
the platform is designed in order to support systems where IoT-based and data intensive applications may pose specific requirements for low latency, restricted available bandwidth, or data locality.

\namep~is built on several Open Source technologies for containerisation and implementations of standards, protocols and services for the IoT.
We present the implementation of the platform and demonstrate it in two different use cases.
\end{abstract}


\section{Introduction}\label{sec:intro}

We present the technical features and the lab deployment of \namep, a lightweight
PaaS platform specialized for edge cloud computing applications and natively designed
for the IoT domain. The general aim of the architecture is to address some of the technical
issues in order to support IoT-based applications, like, e.g., device roaming, low latency, bandwidth,
and power consumption and quest for data locality. The uptake of data-intensive IoT applications is
apparent in several sectors such as industrial manufacturing, oil and gas provisioning, utilities,
transportation, automotive, healthcare, mining, highly remotely distributed contexts (rural,
offshore, logistics) and online games based on augmented reality.

Fog computing \cite{Cisco,Bonomi} is expected to solve core technical issues related to such scenarios.
In fact, such approach is meant to bridge between the IoT and Cloud domains. However, IoT is a natively
distributed paradigm whereas the latter is traditionally a centralised one. Actually, the two approaches
can be integrated as long as two key mandatory challenges are properly addressed, i.e., {\em heterogeneity} of the connected ``things'' and {\em cloud-awareness} of the applications leveraging sensed data.

{\noindent \em Heterogeneity:} IoT sensing devices and communication protocol are using a wide range of technologies, such as RFID, ZigBee or LoRa. Those technologies are used in a large spectrum of application domains, such as e-Health, transport or precision agriculture. Thus, each application domain usually implements its own data models, which further augment heterogeneity issues. In the recent years, several initiatives have been focusing on the development
 of standard support for IoT based applications. Aiming at the integration of different sensing technologies,
 the RIOT platform \cite{riot} has been proposed in order to minimize the hardware
dependent code and to allow for the inclusion of new sensing boards abstracting from the kernel itself.
\namep~complements such an approach by  means of the IoT gateway, which is able to perform as an interface
adapter for multiple technologies, thus easing the access to objects' data from the Internet \cite{Miorandi2012}.
Hence, the problem of accounting for heterogeneous sensor objects is delegated to a specific hardware module.
The IoT gateway in fact can hosts multiple radio-access technologies and it is empowered with dedicated network management modules.

{\noindent \em Cloud-awareness:} cloud-awareness is addressed by \namep~ along two main axes.
Firstly, the platform tackles requirements of scalability, fault tolerance and high availability of computational resources. In fact, it is meant to provide cloud-aware features close to data sources, i.e., granting locality
of computation with respect to data generation. More precisely, moving computation to the cloud edge is done by leveraging on cloud-native patterns in order to match applications' requirements and resources' availability.
Secondly, \namep~manages available network resources in order to process efficiently in case of large amounts of sensed data and connected objects. Moreover, applications performing data crunching may pose specific constraints for the cycle of data consumption and processing and may involve possibly actuation in the loop. In this context, in order to fulfill the requirements of such applications and optimize resources utilization both on the edge cloud and on connected objects, it is crucial to manage the placement and scheduling of workload as a function of existing resources on the edge and on the central cloud.
In \namep~ it is hence possible to perform a scalable deployment of applications reserving edge deployment for those applications in need to be installed closer to the sensing devices. Moreover, the platform will be able to migrate an application to the central cloud (offloading) whenever required, for example, in case of scarce edge resources.

By complementing the IoT domain with features like automatic deployment and dynamic configuration, \namep~allows
the application of the {\em Infrastructure as Code} \cite{infras_ascode} paradigm to the IoT domain itself.
Managing the whole computing infrastructure (the cloud, the edge and the gateway) as a virtualized provision-able infrastructure, combined with the capabilities to package and offer middleware and services for the IoT according to  a micro-service paradigm, allows to address automatic deployment, dynamic configuration and flexible re-provisioning of the whole enabling technology stack. At the gateway, edge and cloud level, infrastructure and IoT services become programmable elements to be combined and orchestrated according to the specific IoT requirements.
While, for example, in the cloud it is possible to address scalability and high availability requirements of an application, at the edge level it is possible to address mission critical real time computing requirements and at the IoT gateway level it is possible to address configurable device protocol support and provisioning, configurable data acquisition and filtering, device-specific business logic delivery.

A short outline of this paper follows: in the next Sec.~\ref{sec:papaya} we describe the architecture
of \namep, in Sec.~\ref{sec:impl} we provide technical insight into the platform and we detail the
specifics of our implementation. In Sec.~\ref{sec:usecases} we describe the use cases and storyboard provided during the demo of the platform. Finally in Sec ~\ref{sec:concl} a recap of the activity carried out and an overview of possible generalizations and enhancements are presented.

\section{Architecture}\label{sec:papaya}
\begin{figure}[t]
\centering
\includegraphics[width=\linewidth]{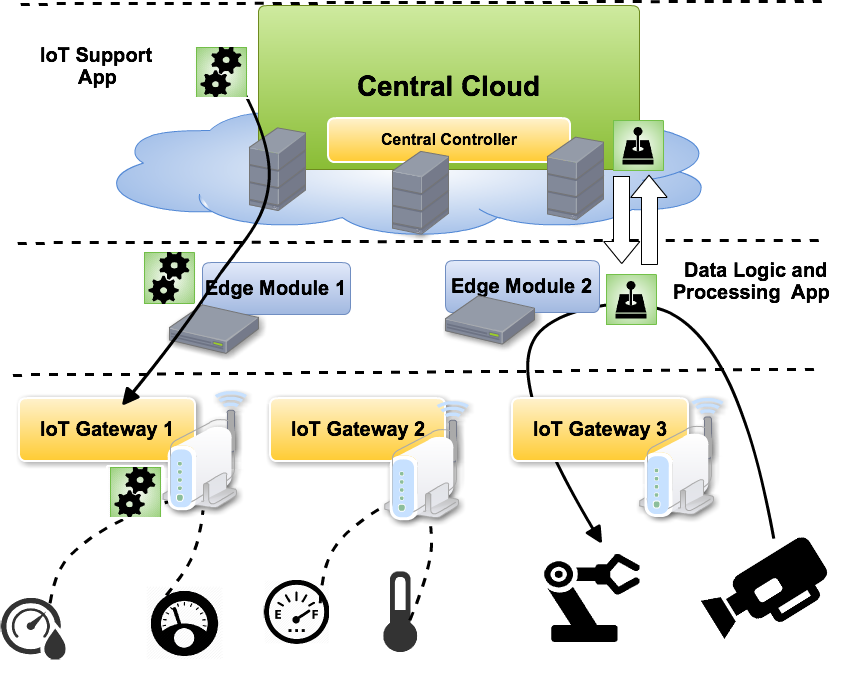}
\vspace*{-7mm}
\caption{The \namep~ logical architecture: (top) central cloud, (middle)
edge modules, orchestrated by the central cloud, (bottom) IoT gateways for
seamless object's integration. On the left: typical deployment of an \Iotap.
On the right, a sample deployment pattern for a \Datap,
to be installed on the edge or on the central cloud at the need.}\label{fig:edgecloud}
\vspace*{-3mm}
\end{figure}˜
The logical Architecture of the \namep~ platform is represented in Fig.~\ref{fig:edgecloud}.
The scheme reports on the three main blocks: a central/traditional cloud platform, an edge-cloud layer and
a set of gateways.

{\em Central cloud platform:}  this cloud hosts the central controller functionality, which assigns distributed resources and provides fleet management services, and offers additional capacity in case of scalability needs. In particular, it performs dynamic configuration of remote resources, e.g., the IoT Gateways and deploys both IoT support applications, e.g. required IoT modules, and/or data logic \& processing applications. The central controller essentially works like an {\em orchestrator and a scheduler} for the workload running on the whole platform, hence offering services similar to the ones provided by a PaaS. In \namep~we have implemented this layer on a private cloud platform, based on OpenStack \cite{openstack}.

{\em Edge Cloud Modules:} these components are designed to let data-intensive applications run close to the IoT Gateways (i.e. where the data are produced). They are small-sized servers with computational power and memory storage capabilities. The central cloud orchestrator employs edge cloud modules in order to migrate applications at the need, so that the workload can be offloaded from the edges to the central cloud and vice versa at the need. Moreover Edge Cloud Modules complement the IoT gateways making the system more resilient (e.g. to network failures) and less dependent by the central cloud thanks to their computational and storage capacity.

{\em IoT Gateways:} gateways represent the hardware interface with objects and are able to handle the multitude of communication technologies and protocol stacks found in the IoT domain. They are hardware platforms ruling the acquisition of the sensor objects' data. They have limited memory and computing capabilities and interface with the edge cloud modules.

\subsection{Supported applications}
There are two main types of applications that are supported by the \namep~ platform and that play a role in the demonstration described in Sec. ~\ref{sec:usecases}:
{\em \Iotap{s}}  and {\em \Datap{s}}.

{\noindent \em \Iotap{s}}: those are service applications specific to the IoT
domain which support the deployment and the maintenance of new objects/sensors in the field.
In particular, these applications are able to:
\begin{enumerate}
\item perform the discovery protocol for new objects attached to the IoT Gateway;
\item retrieve the version of the firmware suitable with the model and the OS of a new object at the need;
\item dispatch the data collected on the edge modules connected to the IoT Gateway;
\item require the installation of new applications in order to manage newly acquired and/or updated objects.
\end{enumerate}

{\noindent \em \Datap{s}}: these applications are deployed, scheduled and orchestrated from the central cloud onto the edge modules according to the current platform condition. Based on the users' latency requirements and/or the amount of data to be processed, such applications can be deployed either to the central cloud or to the edge and then migrated from one to the other. The orchestrator scheduler has been equipped with a simple threshold-based algorithm in order to perform workload management and match application-related constraints; this simple implementation is meant to demonstrate the flexibility of the platform in handling different user's and resources' constraints. We have deferred the implementation of more advanced orchestration logic, e.g., Mesos-like \cite{mesos} schedulers, as part of further versions of \namep.
\subsection{Orchestration}
\namep~ leverages the containerization technology in order to attain automatic deployment, dynamic configuration and orchestration of both \Iotap{s} and \Datap{s}. As detailed in the next section, the platform adopts state of the art Open Source frameworks (like Docker \cite{docker} and Kubernetes \cite{kubernetes}): we have opted for these frameworks because they can be ported onto different hardware platforms. Hence, all layers depicted in Fig.~\ref{fig:edgecloud} are orchestrated in a flexible yet effective way and their required functionalities can complement each other. In the context of today's IoT data-intensive applications, such architectural choice offers a solution able to manage a complex and distributed framework and yet retains and satisfies the main cloud technology features like resilience, robustness, high availability, scalability and elasticity.

\section{Implementation}\label{sec:impl}

We detail hereafter the technical details for each of the modules composing
\namep, with specific reference to the technologies employed in our
installation (see Fig.~\ref{fig:cloud4iot_arch}):

\begin{figure}[t]
\centering
\includegraphics[width=\linewidth]{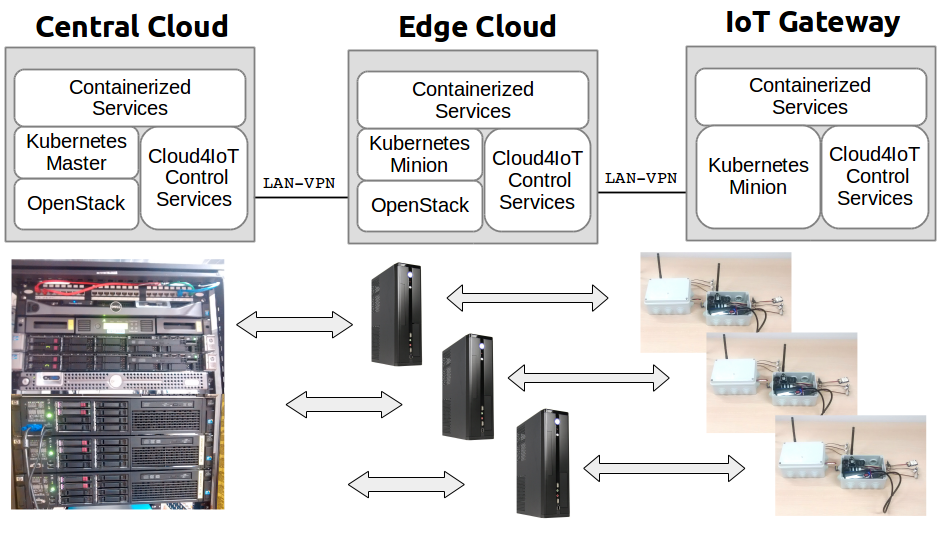}
\vspace*{-7mm}
\caption{The \namep~deployment architecture. On the top a view of services is sketched. On the bottom some pictures of the physical components are provided.}\label{fig:cloud4iot_arch}
\vspace*{-3mm}
\end{figure}˜


{\noindent {$\bullet$} \em Central Cloud.} The Central Cloud runs on dedicated servers,
and offers IaaS service implemented using the Open Source OpenStack platform.
The OpenStack environment, configured for High Availability (HA), is composed
by 3 Controllers nodes, 2 Compute nodes.  The compute servers (HP ProLiant
DL380 Gen9) are equipped with 2xCPU Intel Xeon E5-2630 v3 2.4 GHz
8Cores/16Threads, 96GB RAM, 2x500 GB SATA 7.2K HDD.  The controller servers
(HP ProLiant DL360e Gen8) are equipped with 1xCPU Intel Xeon E5-2407 v2 2.4 GHz
4Cores/4Threads, 32GB RAM, 2x1TB SATA 7.2K HDD. The OpenStack Block Storage
service (cinder) is integrated in OpenStack by means of HP LeftHand iSCSI SAN
and offering 11TB of storage capacity.\\
\indent On top and aside OpenStack we've installed Kubernetes and the  \namep~control services
that take care of the set-up and configuration of the platform and the deployment of the
Kubernetes agents on the Edge Cloud.

{\noindent {$\bullet$} \em Edge Cloud.} Some of the OpenStack IaaS services and agents run
on the Cloud Edge in order to seamlessly deploy the application in the Cloud
Edge Modules. Hence we have extended the OpenStack platform and Kubernetes
installing compute and storage services and the required agents on
the Edge Cloud servers, along with the \namep~ edge control services.
The main purpose of the \namep~control services on the edge layer is to deploy the Kubernetes agents and to provision the IoT Gateway using PXE with a custom image.

The units of the edge cloud installation are based on mini-itx form factor
motherboard.  This allow to fit small space requirements.  In order to test
different computing capabilities and power consumption we have different
CPUs for each unit (from the most capable): Intel
Xeon E3-1226, Intel Core i7-4790S, Intel Avoton C2750. All of them are equipped
with multiple Ethernet network cards, 16GB RAM and 480GB SSD.
On our implementation each edge has one computing unit, with a total of 3 cloud edge nodes.


{\noindent {$\bullet$} \em  IoT Gateway.} The IoT Gateway acts as the network provider for
non-IP IoT devices (e.g., wearables, home/office automation devices that
communicate over Bluetooth Low Energy, ZigBee, Z-Wave, etc.).  It also hosts a
containerised service that provides the essential software components (e.g.,
drivers and protocol implementations) for the connected IoT devices and
application logic (e.g., when and where data from devices should be forwarded).
It is an embedded device with limited resources for data processing and
handling.  It consists of a RaspberryPi version 3 device (1.2 GHz 64-bit
quad-core ARM Cortex-A53 CPU, 1GB RAM, 16GB SD card storage, WiFi/BLE/Ethernet
default connectivity) and an extension shield able to host various
wireless communication modules.


The Edge Cloud provisions the IoT Gateways by installing a
minimal image with \namep~control services able to deploy and configure the Kubernetes agent.


\section{Storyboard and Use Cases}\label{sec:usecases}

In this section we provide the description of a demonstration which has been prepared in order to
showcase the two main features offered by \namep~in its current implementation. The first one is the {\em IoT Roaming} use case. It supports the automatic configuration and re-configuration of IoT Gateways, i.e., \namep~ is able to
react to devices roaming from one IoT Gateway to another. The second feature is the {\em Application Scaling} use case for an \Datap~deployed on the edge layer. Such application can be initially deployed near the sources, scaled at the need within the edge layer, and eventually re-deployed onto the central cloud.

Hereafter we provide the two use cases and the related application of \namep.

\subsection{IoT Roaming: device discovery and automatic service migration}
In this use case we will demonstrate the feasibility of scenarios where sensor objects are moving from one gateway to another. It will show how the roaming can be handled automatically by the platform giving continuity to the service performing sensor data collection. An example of this scenario can be a patient wearing a sensor (smart bracelet) for monitoring life parameters at home. Once the patient leaves his/her home and reaches the hospital, the wearable device is automatically associated with a new gateway and monitoring service is configured accordingly. It must be pointed out that this use case doesn't make any use of 3G/4G mobile networks  or of any smart device (e.g. smartphone) but just a sensor object connected via low level protocols (e.g. bluetooth) with the IoT Gateway; hence it is applicable everywhere roaming of IoT sensors takes place.

The sample IoT roaming demonstration comprises the following steps:\\
{\indent 1)} A wearable device, i.e., a sensor object, is automatically discovered and associated to the IoT Gateway;\\
{\indent 2)} Upon notification of the IoT Gateway, a dedicated \Iotap, orchestrated by the central cloud controller, is deployed in order to manage that object. Simple business logic is provided containerized and customized based on some user's preferences;\\
{\indent 3)}  The sensor object is managed directly by \namep: data collected by the newly added object is acquired by the current IoT Gateway and moved to the edge cloud layer where can be processed in an efficient way;\\
{\indent 4)}  The sensor moves out of range of the current IoT Gateway and associates to a new one;\\
{\indent 5)} The \Iotap~follows the sensor's roaming: together with the current user status, it is automatically deployed and configured onto the new IoT Gateway;\\
{\indent 6)}  Data collected is sent to the edge cloud layer and application status is centrally updated in the cloud.

\subsection{Application Scaling: workload management}

A typical scenario imposing scaling requirements for IoT applications is the Smart City one, where thousands of sensors may be deployed and terabytes of data are potentially collected (e.g., weather, traffic and transportation data, images and videos). Data must hence be processed locally: in fact, bulk transfer to a central cloud is costly and indeed not suitable for real-time usage. Thus, hen IoT objects generate huge amount of data, computational capacity can be conveniently deployed on the edge cloud layer to avoid bulk transfers.

The sample application scaling demonstration comprises the following steps:\\
{\indent 1)} One or more data-intensive applications performing data analytics are deployed on the edge cloud layer;\\
{\indent 2)} Data collected by sensors are moved to the edge cloud layer in order to be elaborated and aggregated;\\
{\indent 3)} Applications get data and elaborate it providing aggregated results (e.g., monitoring data with temporal vs spatial aggregation, log file analysis, shape/object recognition); \\
{\indent 4)} Aggregated results are sent to the central cloud and displayed: the data transfer rate is much lower than the one expected in case of a bulk data transfer;\\
{\indent 5)} Much more data is collected and applications on the edge cloud layer are scaled up in order to serve the new workload;\\
{\indent 6)} Resources on the edge cloud layer are going to be exhausted. The customized scheduler of the orchestrator component offloads the application to other edge cloud modules or to the central cloud. Offloading is performed accounting for the status of the resources in the edge cloud, the available bandwidth, and the requested latency.

\section{Conclusion}\label{sec:concl}
\namep~ provides seamless integration of IoT and Cloud. Architecture patterns typical of Cloud Computing (e.g., Infrastructure as Code) are combined with patterns from IoT. It covers features typical of both approaches and responds to requirements of flexibility, scalability, fault tolerance, and distribution of IoT and data-intensive applications.

The automatic deployment and orchestration of \Iotap{s} and \Datap{s} is performed using containerization solutions based on Open Source technologies. We have demonstrated the potential of this integrated platform in two use cases: IoT Roaming and Application Scaling.

We aim at enhancing \namep~trying to automate and ease installation and operations, and to support different topologies (e.g. generalizing relationships among Central Cloud, Edge Layer and IoT Gateway) and/or to improve workload management. Contribution to the relevant Open Source communities could be also an option.
Doing so, we plan to provide a framework where developers can build new products which fit aforementioned and even novel scenarios.


\bibliographystyle{IEEEtran}
\bibliography{biblio}

\end{document}